**Free Energy Driven Transfer of Charge in Dense Electrochemically Active**

**Monomolecular Films**.

Dmitry Zaslavsky[#], Andrei Pakoulev[¶], Vladimir Burtman[¶*]

University of Illinois, School of Chemical Sciences, 600 South Mathews Street

Urbana, IL, 61801

**Classification:** Physical Sciences, Chemistry.

**Key words:** monolayers, connective networks, in-plain currents, redox centers, cation-radicals, charge transfer, light harvesting.

# Department of Biochemistry

[¶] Department of Chemistry

*Correspondence and requests for materials should be addressed to V. Burtman (current address: University of Utah, Department of Physics; burtman@physics.utah.edu)

The interest in monomolecular films as electric conductors arises from the search for innovative materials. The utility of non-covalently bonded films is limited because they are mechanically unstable and consist of poorly connected domains. Consequently, charge transfers in these films are limited to the distances in the order of a micrometer. Here we show that a recently developed gas phase assembling method (Burtman, V., Zelichenok, A., Yitzchaik, S. (1999) *Angewandte Chemie Inter. Ed.* **38**, 2041-2045.), which produces robust dense monolayers of NTCDI[*] covalently attached to the surface of silicon, allows one to overcome this scale limitation. These virtually insulating monolayers can be photo-chemically populated with cation-radicals via ejection of electrons into the semi-conducting base. The positive charges of cation-radicals can migrate as far as several millimeters within microseconds in a random walk fashion thus demonstrating the macroscopic connectivity of the film. Since the charges exist as cation-radicals, which are potent oxidants, their migration is coupled to transfer of the free energy of their reduction and is driven by the redox potential gradient. Reduction of cation-radicals by an anode converts this free energy into electromotive force. We show how these films can be implemented in solar energy conversion and basic time-resolved distance-controlled studies of sequences of ultra-fast electron transfers.

---

[*] 1,4,5,8-naphthalene tetracarboxylic diimide

The vast majority of studies of charge transfer in monomolecular films was limited to the self-assembling Langmuir-Blodgett monolayers[1,2,3]. These studies have revealed that electrochemical activity of the assembling molecules is a requirement for charge transfer within these monolayers. In other words, electric currents in organic materials represent convolution of elementary redox reactions[4,5,6].

The fundamentals of electron transfers were revealed by analysis[6,7,8,9,10] of electron transfers in structurally defined natural and synthetic molecules. These studies accounted for the role of the intervening medium that couples redox centers, introduced σ-coupling (via covalent or hydrogen bonds) and π-coupling (via overlapping π-orbital) and through-space jumps and thus refined the empiric rule of the exponential drop in the tunneling rate constant as the width of the separating barrier increases.

In biological electron-transfer complexes[8,9] and redox-active polymers[3-5] the exponentiality problem is overcome by switching to the multi-step transport through a continuous network of σ-coupled redox centers. Even though the σ-coupling is efficient within macro-chains[11] or thick films[12], the ways of creating extended continuous σ-coupled 2D networks are yet to be found. Despite that π- and through space couplings provide smaller range for the single-barrier tunneling[13] than the σ-coupling[9,10], the multi-step electron transfer through the π-stacked purine strands of DNA has proved itself efficient[6,14,15].

The non-covalently bonded monomolecular films can provide connectivity between redox centers within grains of typically a micrometer or even smaller[3]. Extension of the network into the macro scale requires stabilization of the structure, preferably by covalent bonding, making the network rigid. At the same time, the number of defects introduced into the structure during the assembly has to be kept minimal in order to avoid formation of isolated domains. Therefore, the electrochemically active "building" elements have to be small in order to ease their packing. Advantageously, macroscopic π-stacked 2D networks of small aromatic diimides can be assembled on solid surfaces by the recently developed method[16] utilized in this work (**Figure 1A**). We show that these monomolecular films can sustain charge transfer over at least several millimeters. This long-range charge transfer originates from and testifies for the macroscopic connectivity of the fabricated films. This connectivity can be used in many applications including solar cells and photomultipliers.

**METHODS.**

**Preparation of organic films.** Thick, impermeable to light, wafers of n-type (500 Ω cm, Virginia Semiconductors) Si(100) having 20 Å of $SiO_2$ coating and glass slides were cleaned and functionalized with amino groups (**step a**) as described in 16. NTCDA[†] evaporated at 110˚C and reacted with the $NH_2$-functionalized surface for 45 min in a Bell Jarr chamber at $10^{-5}$ Torr (**step b**). The product of assembly lacks the amino group preventing formation of the second layer. The substrate was kept on heated sample holder (180 ˚C) preventing phys-adsorption of the precursor. Vacuum deposition modified over 60 % of the surface amino groups[16,17]. A layer of 4-aminophenylthiol was added within 20 min (**step c**). This reaction tops the surface with SH-groups reactive towards metals[18]. The chip was gently rinsed with isopropanol and heated at 80 °C for 1 hour. A silver contact was placed on top of films by placing a drop of colloidal silver in acetonitrile and allowing the solvent to evaporate, the area or the contact between silver and coated silicone was ~ 3 mm$^2$. Thick films of NTCDA and C6-NTCDI were prepared by phys-adsorption of NTCDA on the cold functionalized surface of silicon. Sparse NTCDI films were obtained from thick films of NTCDA by heat de-sorption of the excess of NTCDA followed by **step c**.

---

[†] 1,4,5,8-naphthalene tetracarboxylic anhydride

**Determination of the absorption and external quantum efficiencies.** The spectra of dense and sparse monolayers of NTCDI grown on glass slides were recorded in a Shimadzu spectrophotometer. Comparison of the films grown on Si and glass is justified by the surface titration, which shows that the density of amino-groups on glass and SiO$_2$ coating of Si(100) is approximately the same[17] (2-3 per 100 Å$^2$). The EQE was calculated by referring the current generated by the photo-element through a 500 Ω resistor to the output of a Xenon lamp/ monochromator (8 nm band) standardized to the calibrated light source from Oriel. It was also determined at 532 nm by referring to the non-saturating output of a CW YAG laser calibrated with a photo-calorimeter.

**Time-resolved photo-voltage measurements.** These were performed with a picosecond laser system consisted of a Ti:Sapphire femtosecond laser, stretcher/compressor/amplifier of picosecond laser pulse ("Titan", Quantronix) and optical parametric amplifier of superflourescense ("TOPAS", Qunantronix/Light Conversion). The output pulse parameters were: wavelength 532 nm, energy 10 μJ, pulse duration 1 ps, and repetition rate 1 KHz. To avoid saturation effects the pulse energy was attenuated to a 0.1 μJ level with a beam diameter of 1 mm. The photo-voltage transients were recorded by Tektronix TDS 3025 digital oscilloscope by-passed with a 7.6 kΩ resistor.

**RESULTS and DISCUSSION.**

We topped the film with a small-area silver electrode and found that the resulting Ag/NTCDI monolayer/Si sandwich (**Figure 1B**) was sensitive to light. In the dark the current-voltage (I-V) curve goes through the origin (**Figure 2**) and straightens at the slopes that correspond to ~ 30 and ~ 10 kΩ respectively as the positive or negative voltage increases. Illumination with continuous monochromatic light changes this characteristic dramatically. Now the I-V curve also goes through the bottom right quadrant where the current through the element opposes the external bias. This plainly manifests that light generates electro-motive force, which induces a negative charge on silicon (photo-cathode) and positive charge on silver (photo-anode).

The mutual arrangement of the cell and the incident light; the external quantum efficiency (EQE) of 0.4-0.7 (**Figure 3**), which cannot be accounted for by the absorption efficiency of the film; the dissimilarity of the EQE and the absorption efficiency of the film and resemblance of EQE to the spectrum of silicon[19] suggest that the light productively absorbs by the semiconductor (**Figure 1B**). This contrasts with light-harvesting in the dye-sensitized photo-voltaic cells[20,21] or the hybrid nanorod-polymer cells[22], which require special junctions with enormous contact areas between the materials to enhance their absorption efficiency. The geometry of the cell also suggests that light harvesting incorporates a longitudinal spatial energy transfer.

In contrast to natural photosynthesis[23], the conserved energy does not migrate in the form of excitation. The absence of long-range light harvesting, when the uncoated side of the chip was illuminated, establishes importance of the film and rules out energy transfer within the silicon bulk. The process starts with a redox reaction separating charges between silicon and film (**reaction 1**). Due to the spatial separation of the cathode and the anode chemistry, this reaction can be identified straight from the polarity of the cell, which shows that electrons from the NTCDI molecules are ejected into silicon, *i.e.* NTCDI molecules oxidize. Charge separation is supported by the asymmetry of multiplication of the dark current with respect to the bias direction (**Figure 2**), which cannot be accounted for by the influence of non-polar excited states. Indeed, regardless of the amplification mechanism, they would not be able to discriminate the bias directions. On the contrary, charge separation explains this asymmetry easily. Since the positive charges in the NTCDI film "reflect" in silicon "mirror", they are accompanied by their counter-charged electrostatic images and migrate as trans-surface dipoles. As these dipoles approach the Ag/film/Si junction, they are either attracted into the contact area by parallel or repelled by the anti-parallel external field. In agreement with photo-oxidation mechanism of energy conservation, light harvesting was observed in the thick films of N,N'-dihexyl-naphthalene tetracarboxylic diimide ($C_6$-NTCDI) phys-adsorbed on the silicon surface, but not in the thick films of NTCDA, because NTCDA is much harder to oxidize.

This photo-oxidation incorporates several events. Absorption of light results in elevation of an electron to the conductance band leaving a vacancy in the valence band upon. This vacancy can be filled either by back recombination of the electron from the conductance band (unproductive decay) or by transfer of an electron from the film into silicon (**reaction 1**) since the affinity of this vacancy for an electron is apparently high enough to oxidize a molecule of NTCDI. As a result, similarly to the photochemistry of dye-sensitized cells[20,21], an extra electron remains in the conductance band of silicon, while the oxidized molecules of NTCDI become cation-radicals (**Figure 1B, inset**). These radicals resemble those in protein[24,25] and DNA[26] molecules and combine essential structural motifs of the oxidized forms of two ubiquitous redox cofactors: $NAD^+$ and semiquinone. These features ease oxidation of neutral molecules.

The energy of light is conserved in two forms: the electrostatic energy of trans-surface dipoles and the free energy of reduction of cation-radicals by silver (**reaction 3**). The equilibrium of this reaction is shifted towards neutralization of cation-radicals and we could not attain electric currents in the films assembled on the surface of glass, because measurements of such currents even over micro gaps require very sensitive equipment[27]. To study migration of cation-radicals to the photo-anode (**reaction 2**), we used short light pulses focused at different distances from the anode to provide both time and spatial resolution. The photo-voltage transients consist of voltage rise and decay (**Figure 4**). The decay is exponential with $\tau \sim 40$ μs at the given external load. This component represents the discharge of the cell due to both the back-flow and the functional current through the load.

The voltage rise originates from accumulation of charges at the anode and its kinetics depends on the distance between the illuminated spot and silver (**Figure 4**). To avoid a model bias we described the kinetics of the voltage rise by its half-time $T_{1/2}$ (**Figure 4, inset**), which reflects the travel-time of the cation-radicals. Despite some spatial uncertainty due to the finite beam size, the travel-time appears to be proportional to the second power of the distance between the illuminated area and the silver electrode. Since the area (distance quadrate) covered by a random-walking particle is proportional to its travel-time, this dependency suggests that the charges associated with cation-radicals migrate (**Figure 1B, reaction 2**) through the film in the random walk fashion. This migration is a set of successive self-exchange reactions between the neutral molecules and the cation-radicals. It is driven by the spatial gradients of their electro-chemical potential, which is present because the anode reduces the cation-radicals. This migration requires proximity of the molecules forming the monolayer and acting as redox centers, which manifests itself as a broad absorbance band in the green-orange region (**Figure 3**). The migration was not observed in the sparse monolayers of NTCDI.

The slight decrease of the total charge collected by the anode is probably due to partial recombination of trans-surface dipoles. However, the life-time of cation-radicals is apparently somewhat longer than the travel-times of ~1 μs or less at the given distances. The similar back recombination in the dye-sensitized $TiO_2$ takes milliseconds[28].

The charges formed in the remote illuminated area have to travel within the film over "gigantic" distances at finite speed. Therefore, the beginning of the voltage rise should lag behind the initial rapid charge separation. The small initial signal (**Figure 4**), which was also present in the sparse films and was especially strong when the edge of Ag anode was illuminated, allowed only qualitative studies of this delay. However, when the beam was focused 10 mm away from the anode, a ~100 ns lag became visible, providing a rough estimate for the velocity of migrating charges in the order of $~10^5$ m/s (~1 fs/Å). This lag has never been observed in the molecules of DNA where cation-radicals can migrate over ~100-200 Å, depending on the composition of these molecules[6,13,15]. Under most experimental conditions the overall transfer over these distances is limited by the radical formation[29] hiding the related lag, expected to be in the order of ~100 fs.

The efficient delivery of positive charges to the photo-anode becomes possible due to the macroscopic connectivity of the network of NTCDI molecules acting as redox centers. This connectivity permits rapid transfer of large net charge through an extremely narrow cross-section and originates from assembling NTCDI molecules at the positions predetermined by arrangement of the $NH_2$-groups within the dense interlinked siloxane network*[30]*. These networks, stabilized by bonding to solid surface, can be used for coupling electrodes in detectors or organic solar cells, which will utilize semiconductors with high absorption efficiency as the light harvesting antennas. Their configuration is reverse to that of the dye-sensitized cells[20,21] and their components act as photo-oxidant, charge-energy relay and fuel.

**Acknowledgements**: We thank Prof. A. Gewirth, Prof. R. Gennis for their generous support of this work, Prof. D. Dlott, Dr. R. Hash, Dr. R. Strange for the use of their instrumentation. Writing of the manuscript was greatly facilitated by discussions with Prof A. Gewirth, Prof. R. Gennis, Prof. H. B. Gray, Dr. A. Yakimov and Dr. D. Rykunov. We thank E. Kiselov for graphic design and Dr. P. Tsatsos for proofreading the manuscript.

**Legends:**

**Figure 1. Fabrication and operation of the photo-element.**
**(A)** The 3-step assembling of monomolecular films of NTCDI. **(B)** Incident light creates holes (**h**) in the valence band of silicon. Upon ejection of electrons (**reaction 1**) from the immobilized molecules into the holes cation-radicals are formed (**inset**). Cation-radicals rapidly exchange with the neutral molecules (**2**); the positive charge travels through the film, resulting in the **in-plain** current. Re-reduction of cation-radicals by silver (**3**) completes the photo-voltaic element. Reactions **1-3** generate electromotive force. The diode (**1**) reflects rapid photo-ejection of electrons from the film and slow back-reaction. The external electric connections used in the voltammetric experiments (**Figure 2**) include a power source, ammeter, and voltmeter.

**Figure 2. Electro-optical properties of the Si/NTCDI/Ag heterostructure.**
All experiments have been carried out at room temperature. Constant monochromatic 400 nm light illuminated a ~10 mm spot centered ~5 mm away from the silver electrode. When the Ag electrode is highly electronegative, the element behaves as a photo-amplifier and light noticeably increases the asymmetry of the I-V curve.

**Figure 3. Effect of the surface density of NTCDI on the film spectra and light harvesting efficiency.**
Dense film **A** assembled on glass has a prominent aggregation band in the greenish-orange region. The external quantum efficiency of the corresponding Si cell illuminated by constant non-saturating light is represented by spectrum **C**. Sparse film **B** has no aggregation band in the spectrum. The external quantum efficiency of the corresponding cell was to low to measure.

**Figure 4. Scanning time-resolved photo-voltage probing.**
The transients were induced by pulses centered at different distances apart from the silver electrode. The **inset** shows the dependency of the $T_{1/2}$ of the voltage rise on the distance $l$ between the bright spot and the silver photo-anode. The solid line approximates the points by a parabola $T_{1/2} = l^2/D$ with the parameter $D \sim 10^6$ cm$^2$/s.

A

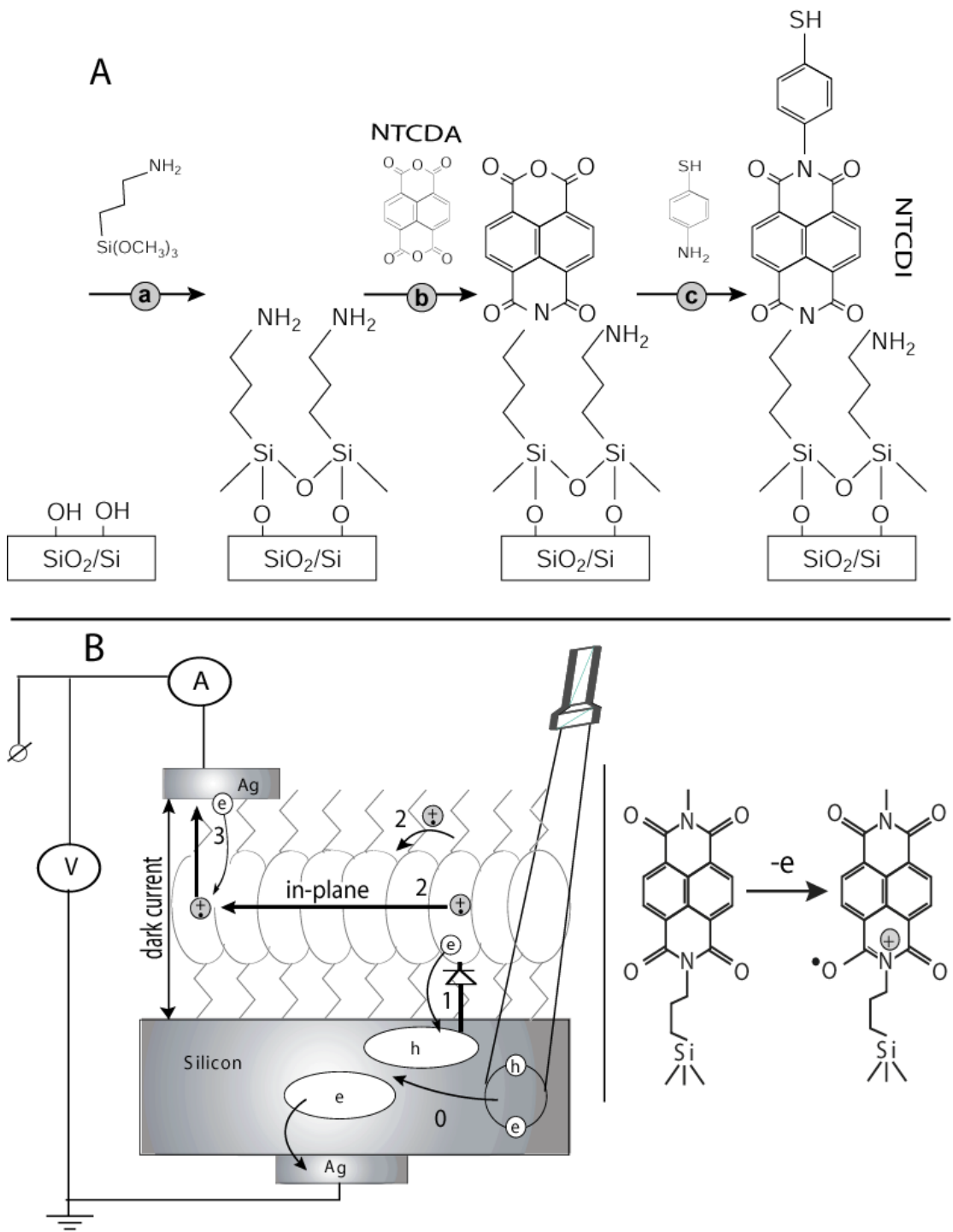

B

Fig 1

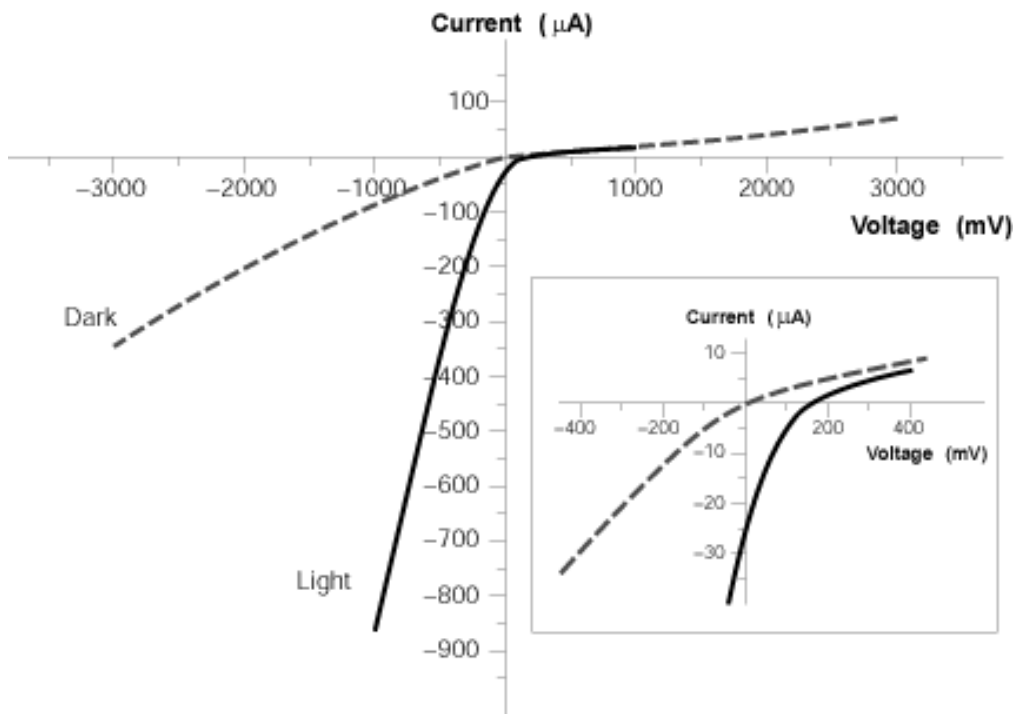

Fig 2.

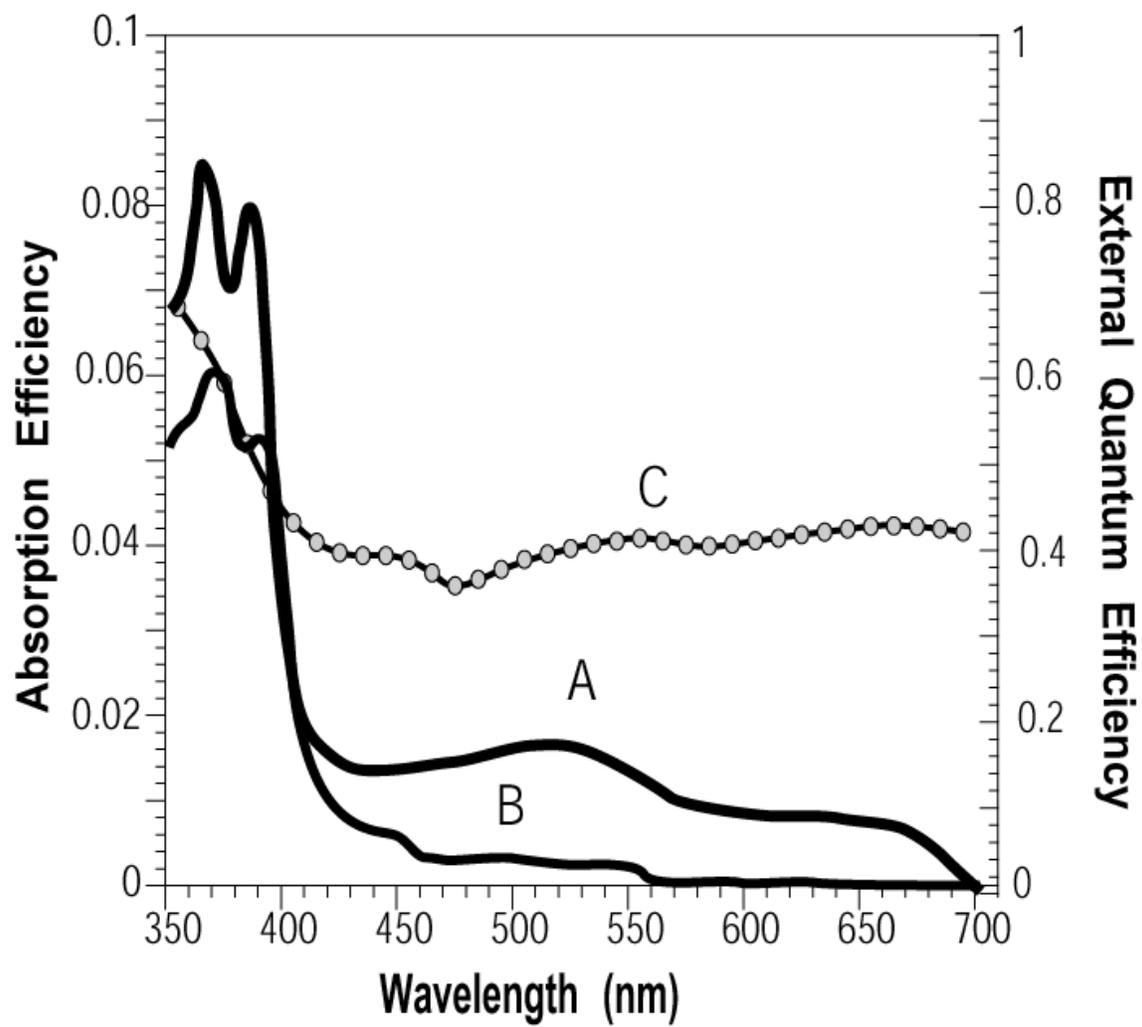

Fig 3.

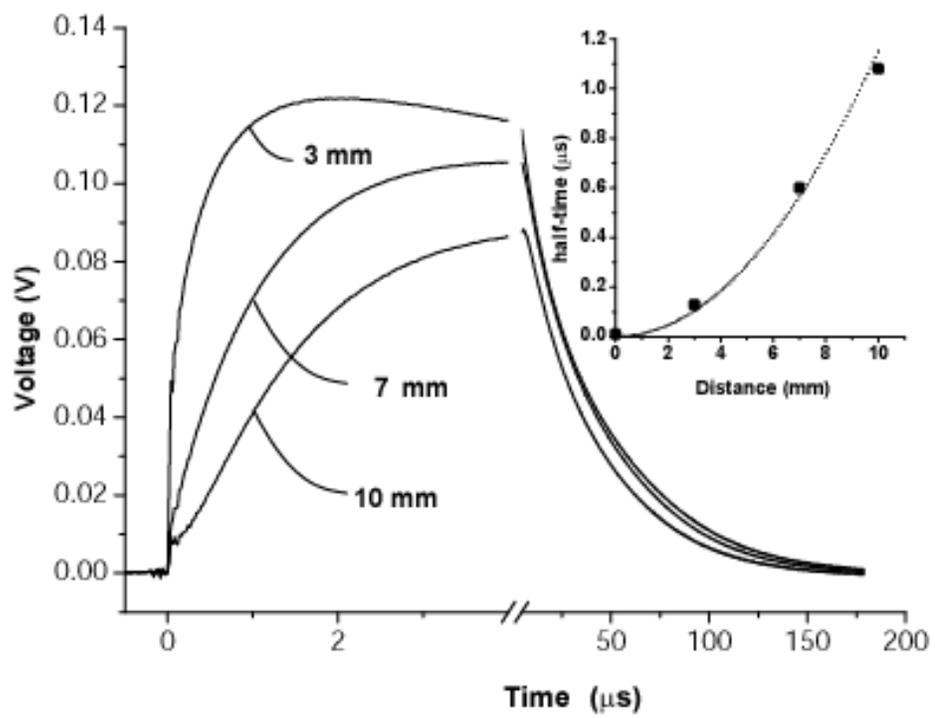

Fig. 4

1. Bryce, M R., & Petty M. C. (1995) *Nature (London)* **374**, 771-776

2. Bjørnholm, T., Hassenkam, T., & Reitzel, N. (1999) *J. Mater. Chem.* **9**, 1975-1990

3. Bernier, P., Lefrant, S., & Bidan G. (ed.) *Advances in Synthetic Metals Twenty Years of Progress in Science and Technology*. (Elsevier Science, 1999).

4. Greenham, N. C., & Friend R.H. (1995) *Solid State Physics* **49**, *Advances in Research and Application*, edited by H. Ehrenreich and F. Spaepen (Academic Press, New York, 1995), 2-150.

5. Novák, P., Müller, K., Santhanam, K. S. V. & Haas O. (1997) *Chem. Rev.* **97**, 207-281.

6. Heath, J. R., & Ratner, M. A. (2003) *Physics Today* **56 (5)** 43-49.

7. Markus, R. A., Sutin, N. (1985) *Biochim. Biophys. Acta* **811**, 265-322.

8．Moser, C. C., Keske, J. M.,  Warncke, K.,  Farid, R. S., &  Dutton, P. L. (1992) *Nature* **355**, 796-802.

9. Winkler, J. R., Di Bilio, A. J., Farrow, N. A., Richards, J. H. & Gray, H. B. (1999) *Pure Appl. Chem.* **71**, 1753-1764.

10. Onuchik, J. N., Beratan, D. N., Winkler, J. R., & Gray H. B. (1992) *Annu. Rev. Biophys.  Biomol. Struct*, **21**, 349-377.

11. Georgiadis, R., Peterlinz, K. A., Rahn, J. R., Peterson, A. W., & Grassi, J. H. (2000) *Langmuir* **16**, 6759-6762.

12. Wamser C.C., Bard, R.R., Senthilathipan, V., Anderson, V. C., Yates, J. A., Lonsdale, H. K., Rayfield, G. W, Friesen, D.T., Lorenz,  D. A., Stangle, G. C., et al. (1989) *J. Am. Chem. Soc.* **111**, 8485-8491.

13. Liu, C.- S., & Schuster, G. B.  (2003) *J. Am. Chem. Soc.* **125**, 6098-6102.

14. Porath, D., Bezryadin, A., de Vries, S., & Dekker, C. (2000) *Nature* **403**, 635-638.

15. Berlin, Y. A., Burin, A. L., & Ratner, M. A. (2001) *J. Am. Chem. Soc.* **123**, 260-268.

16. Burtman, V., Zelichenok, A., Yitzchaik, S. (1999) *Angewandte Chemie Inter. Ed.* **38**, 2041-2045.

17. Burtman, V., Offir, Y., Yitzhaik, S. (2000) *Langmuir* **17**, 2137-2142.

18. Whitesides, G. M., & Laibinis, P. E. (1990) *Langmuir*, **6**, 87-96.

19. S. Sze, *Semiconductor Devices Physics and Technology* (Wiley: New York, 1985).

20. Hagfeldt, A. & Grätzel, M. (2000) *Acc. Chem. Res.* **33**, 269-277.

21. Grätzel, M. (2001) *Nature* **414**, 338-344.

22. Huynh, W. U., Dittmer, J. J., & Alivisatos, A. P. (2002) *Science* **295**, 2425-2427.

23. Glazer, A. N. (1983) *Ann. Rev. Biochem.* **52**, 125-157.

24. Di Bilio, A. J., Crane, B. R., Wehbi, W. A., Kiser, C. N., Abu-Omar, M. M., Carlos, R. M., Richards, J. H., Winkler, J. R. & Gray H. B. (2001) *J. Am. Chem. Soc.* **123**, 3181-3182.

25. Barry, B. A., El-Deeb, M. K., Sandusky P. O., & Babcock, G. T. (1990) *J. Biol. Chem.* **265**, 20139-20143.

26. Kasai, H., Yamaizumi, Z., Berger, M., Cadet J. (1992) *J. Am. Chem. Soc* **114**, 9692-9694.

27. Collet, J., Lenfant, S., Vuillaume, D., Bouloussa, O., Rondelez, F., Gay, J. M., Kham K., & Chevrot C. (2000) Appl. Phys. Lett. **76**,1339-1341.

28. Rehm, J. M., McLendon G. M., Nagasawa Y., Yoshihara K., Moser J. & Gratzel M., (1996) *J. Phys. Chem.* **100**, 9577-9578.

29. Wan, C., Fiebig, T., Schiemann, O., Barton, J. K., & Zewail, A. H. (2000) *Proc. Natl. Acad. Sci. USA* **97**, 14052-14055.

30. Ulman, A. (1996) *Chem. Rev.* **96**, 1533-1554.